\documentstyle[prb,aps,tighten,multicol,epsf]{revtex} 

\begin{document} 
 
\draft 
 
\title{Electron transport through a mesoscopic metal-CDW-metal
junction}
 
\author{I. V. Krive and A. S. Rozhavsky\cite{deceased}}
\address{B.I. Verkin Institute for Low Temperature Physics and
Engineering, Kharkov, Ukraine}

\author{E. R. Mucciolo and L. E. Oxman}
\address{Departamento de F\'{\i}sica, Pontif\'{\i}cia Universidade
Cat\'olica do Rio de Janeiro,\\ Caixa Postal 38071, 22452-970 Rio de
Janeiro, Brazil}

\date{\today} 

\maketitle 
 

\begin{abstract}
In this work we study the transport properties of a finite
Peierls-Fr\"ohlich dielectric with a charge density wave of the
commensurate type. We show that at low temperatures this problem can
be mapped onto a problem of fractional charge transport through a
finite-length correlated dielectric, recently studied by Ponomarenko
and Nagaosa [Phys. Rev. Lett {\bf 81}, 2304 (1998)]. The temperature
dependence of conductance of the charge density wave junction is
presented for a wide range of temperatures.
\end{abstract} 

 
 

\begin{multicols}{2} 
 

\section{Introduction}
\label{sec:I}

The problem of charge transport through an inhomogeneous system (e.g.,
a dielectric junction between two metallic leads) continues to attract
considerable attention from both purely theoretical and practical
points of view. It is especially interesting in the case of a
one-dimensional (1D) junction where the interaction (electron-electron
or electron-phonon) drastically changes the system properties. The
most studied example is a 1D quantum metallic wire connected to 2D
electron reservoirs. It is known that the low energy properties of
electrons in the wire can be described by the Luttinger liquid (LL)
model.\cite{Haldane81} Since electrons do not propagate freely in a
LL, a nontrivial problem arises: How are conduction electrons
converted into propagating charged excitations at the metal-LL
boundary?

A few years ago, it was shown \cite{finiteLL} that, for adiabatic
contacts (i.e., when all the inhomogeneities are smooth on a scale of
the electron Fermi length), the dc conductance of a pure quantum wire
is perfect, $G_0 = 2e^2/h$, and temperature independent (the factor of
two accounts for spin degeneracy). This result is not surprising since
electrons coming from the leads are not backscattered by the effective
barrier represented by the LL piece of the wire. The linear
conductance, according to the Landauer-B\"uttiker
approach,\cite{Landauer} should then be equal to the conductance
quantum $G_0$. Therefore, the dc conductance of a pure quantum wire in
adiabatic contact with source and drain leads is not renormalized by
the interaction.

The electron-electron interaction does affect the current-voltage
characteristics of a LL junction if either impurities are present
\cite{Kane92,Furusaki96} or electrons are backscattered by Umklapp
(U)-processes. U-processes lead, as a rule, to the formation of a gap
in the spectrum of charged excitations and in this case one may
consider the junction as a {\it dielectric} wire.

Recently, Ponomarenko and Nagaosa \cite{Ponomarenko98a} studied
electron transport through a {\it finite} Mott-Hubbard dielectric. By
mapping the problem at low temperatures onto the quantum impurity
problem in an {\it infinite} LL (the so-called Kane-Fisher problem
\cite{Kane92}) they obtained a surprising result: The conductance of a
finite-length correlated dielectric, adiabatically connected to
metallic leads equals, at $T=0$, the conductance of a perfect metal
junction $G_0$. However, an exponentially small conductivity, typical
of a tunnel junction, is restored already at very small temperatures
$T_i \sim \Delta_L \exp(-2\Delta/\Delta_L)$, where $\Delta$ is the
Mott-Hubbard gap and $\Delta_L = \hbar v_\rho/L$ ($L$ is the length of
dielectric and $v_\rho$ is the velocity of the charged
excitations). This result was later generalized \cite{Ponomarenko98b}
to the transport of fractional charge through a finite-length
correlated dielectric.

The purpose of the present paper is to consider the electron transport
through a junction formed by a Peierls-Fr\"ohlich dielectric where a
charge density wave (CDW) is present. The problem of conversion of
electronic current into CDW current at the boundary metal-Peierls
dielectric was studied for the first time over a decade
ago.\cite{Kasatkin84,Krive87} In Ref.~\onlinecite{Kasatkin84} (see
also Ref.~\onlinecite{Visscher96}) the Peierls dielectric was treated
in the so-called adiabatic approximation, when the dynamics of the
electrons is considered for a fixed phonon field (the order
parameter), which is found self-consistently (for a review, see
Refs.~\onlinecite{Gruner94} and \onlinecite{Krive86}). For an order
parameter with a time-independent phase, the transport problem can be
reduced to solving the Bogoliubov-de Gennes equation for the electron
wave function in the piecewise constant effective field of the order
parameter. The calculations are very close to those for a
metal-superconductor junction. There is only one important difference:
While in a superconductor the condensate wave function is formed by
two particles with opposite momenta, in a Peierls state the
corresponding spinor consists of a particle and a hole with momenta
differing by $ 2k_F$. As a result, the scattering of particles on the
off-diagonal barrier produced by the order parameter is not of the
Andreev type.\cite{Artemenko97} For electrons in the metallic leads
with energies inside the gap there is always a strong
backscattering. It is then evident that the quasiparticle contribution
to the conduction will be exponentially small,
$G\propto\exp(-2\Delta_0/\Delta_L)$, for temperatures and voltages
less than $\Delta_0$ ($\Delta_0$ is the Peierls gap at
$T=0$).\cite{Kasatkin84}

One should not forget, however, that there exists another channel for
charge transport in Peierls-Fr\"{o}hlich systems. Namely, the
electronic current coming from the metallic leads can be converted
into a CDW current which freely propagates through the junction. The
process of conversion is perfect when it involves an incommensurate
CDW and the contacts between the Peierls dielectric and the metallic
leads are adiabatic.\cite{Oxman99} In other words, when the phase
degree of freedom of the order parameter in the Peierls dielectric is
correctly treated, one finds that the conductance of an impurity-free
adiabatic CDW junction is perfect at low temperatures. This result was
proved in Refs.~\onlinecite{Oxman99} and \onlinecite{Rejaei} using
different approaches.

In Ref.~\onlinecite{Oxman99} we related the property of perfect
conductance of an incommensurate CDW junction to the existence of an
anomalous chiral symmetry in the system. In this paper we focus our
attention instead on the commensurate case, when the electron filling
factor is a rational number and no chiral symmetry exists. We show
that quantum fluctuations make the conductance of a commensurate,
finite Peierls dielectric perfect at $T=0$, much in analogy to the
case of a Mott-Hubbard dielectric.\cite{Ponomarenko98a,Ponomarenko98b}
As temperature is increased, however, the conductance rapidly
decreases (we call this the instanton regime of quantum
transport). The perfect conductance is lost at temperatures $T \sim
\Delta_i \simeq \sqrt{ \Delta_{L}^{\scriptsize{\mbox{CDW}}}
E_s^{\ast}} \exp(-E_s^{\ast}/ \Delta_L)$, where
$\Delta_L^{\scriptsize{\mbox{CDW}}} = \hbar v_c/L$, $v_c$ is the
velocity of the CDW, and $E_s^\ast $ is the rest energy of quantum CDW
solitons. For higher temperatures, quantum phase fluctuations are
suppressed and the conductance scales approximately as $T^{-2}$,
reaching a global minimum value $G_i\simeq G_0 \exp(-2E_s^{\ast}/
\Delta_L^{\scriptsize{\mbox{CDW}}})$ at temperatures $T \sim
\sqrt{\Delta_{L}^{\scriptsize{\mbox{CDW}}}
E_s^{\ast}}\gg\Delta_L^{\scriptsize{\mbox{CDW}}}$. A further increase
of temperature induces the appearance of fractionally charged solitons
which dominate the conduction in the system (we call this the soliton
regime of quantum transport). Thus, at temperatures $T\simeq
E_s^{\ast}$ the conductance sharply increases, saturating at the value
$G_M = G_0/M^2$ ($M$ is the commensurability index), which also
corresponds to a perfect transport, but of fractionally charged
solitons ($q_s=e/M$) instead of bare electrons. This step-like
behavior in $G(T)$ terminates at $T \sim \Delta_0$, where the
conductance rapidly reaches the pure metallic value $G_0$.

It is important to remark that the Mott-Hubbard and the
Peierls-Fr\"{o}hlich systems are not identical. There are two energy
scales in the latter which are not present in the former. These scales
are the phonon energy $\hbar\bar{\omega}$ (here $\bar{\omega}\equiv
\omega(2k_F)$ denotes the frequency of the phonons with momentum $\pm
2k_F$) and the rest energy $E_s^{\ast}$ of quantum solitons of the
commensurate CDW. In principle, the existence of these additional
scales could lead to a more complicated temperature dependence of the
conductance than that found in Refs.~\onlinecite{Ponomarenko98a} and
\onlinecite{Ponomarenko98b}. However, we will argue that the phonon
energy scale, in particular, does not lead to any new features.

This paper is organized as follows. In Sec.~\ref{sec:II} we introduce
the effective phase Lagrangian that describes commensurate CDW
junction. The conductance and its temperature dependence in the
instanton regime are discussed in Secs.~\ref{sec:III}; the soliton
regime is explored in Sec.~\ref{sec:IV}. Final remarks and the
conclusions are presented in Sec.~\ref{sec:V}.

\section{The Peierls-Fr\"ohlich system}
\label{sec:II}

The low-energy dynamics of a one-dimensional electron-phonon system is
normally governed by the Lagrangian\cite{Krive86}
\begin{eqnarray}~
{\cal L} & = & - \frac{\Delta^2}{g^2} + \frac{\dot{\Delta}^2 +
\Delta^2\dot{\varphi}^2}{g^2\bar{\omega}^2} + \bar{\psi} \left( i\hbar
\gamma_{\mu}\partial^{\mu} - \Delta e^{i\gamma_5\varphi} \right) \psi
\nonumber \\ & & + {\cal L}_{\scriptsize{\mbox{com}}}.
\label{eq:1}
\end{eqnarray}
Here, $u(x,t) = (\Delta/2) \cos(2k_F x +\varphi)$ is the magnitude of
the phonon field with momentum $Q \simeq \pm 2k_F$ and $g$ is the bare
constant of the electron-phonon interaction. We have defined
$\bar{\psi} \equiv \psi^{\dagger} \gamma_0$, with $\psi^{\dagger} =
(\psi_L^{\ast},\psi_R^{\ast})$ comprising left- and right-moving
electron fields. The Dirac matrices $\gamma_{\mu}$ ($\mu=0,1$) and
$\gamma_5$ in 1D are represented by the Pauli matrices, whereas the
partial derivatives follow the convention $\partial_{\mu} \equiv
(\partial_t, v_F\partial_x)$. The last term in Eq.~(\ref{eq:1})
represents the commensurability energy \cite{Krive86,Lee74} that
exists for systems with rational electron filling $\nu = N/M$ with
respect to the underlying lattice model ($N<M$). The commensurability
index is defined as $M = \pi N/k_Fa > 2$, where $a$ is the lattice
constant. The appearance of the commensurability energy is due to
$M$-fold Umklapp scattering. This term will be specified later.

The interacting 1D electron-phonon system model presented in
Eq.~(\ref{eq:1}) has already been studied by many authors (e.g., see
reviews in Refs.~\onlinecite{Krive86} and \onlinecite{Fukuyama85}). It
was shown that a gap $\Delta_0$ develops in the electron spectrum at
low temperatures. For spin-1/2 fermions and for a vanishingly small
phonon frequency $\bar{\omega}\rightarrow 0$, even a weak
electron-phonon interaction produces a gap. Let us, for simplicity,
consider only spinless electrons. In this case, for a finite
$\bar{\omega}$, the electron-phonon interaction should be sufficient
to produce a gap,\cite{Voit87} while, at the same time, the junction
can be sufficiently long to make the influence of finite-size effects
\cite{Krive83} on the gap negligible. In the classical Peierls theory,
where all fluctuations are neglected, the gap for a weak
electron-phonon coupling ($\lambda = 2g^2/\pi\hbar v_F\ll 1$) takes
the familiar BCS-like form $\Delta_0 \propto
\varepsilon_F\exp(-1/\lambda)$, where $\varepsilon_F$ is the electron
Fermi energy. For a quasi-1D systems, this gap is the modulus of the
order parameter $\Delta_0e^{i\varphi}$ of the Peierls state which
develops at temperatures $T<T_P$ (the Peierls transition temperature
$T_P$ is determined by the strength of inter-chain interaction and,
usually, $T_P\ll \Delta_0$). The fluctuations of the phase field
$\varphi$ in a pure 1D system destroy the long-range
order. Nevertheless, the fact that $\Delta_0 \neq 0$ minimizes the
energy of the system shows that at $T\ll \Delta_0$ the fluctuations of
the modulus of the phonon field are strongly suppressed.

In what follows we will assume that the ground state of a sufficiently
long ($L \gg \hbar v_F/\Delta_0$) CDW junction is characterized by a
gap $\Delta_0$ in the single-particle spectrum. This gap will be the
largest relevant energy in the problem
($\hbar\bar{\omega}\ll\Delta_0$).

To proceed further, it is convenient to bosonize the fermion degrees
of freedom in Eq.~(\ref{eq:1}): according to standard
rules\cite{Coleman75,Mandelstam75},
\begin{equation}
i\bar{\psi} \gamma_\mu \partial^\mu \psi \Longleftrightarrow
\frac{1}{8\pi v_F} \partial_\mu \eta \partial^\mu \eta,
\label{eq:2}
\end{equation}
\begin{equation}
\bar{\psi} \psi \Longleftrightarrow \frac{1}{\pi\alpha} \cos\eta,
\label{eq:3}
\end{equation}
and
\begin{equation}
\bar{\psi} \gamma_5 \psi \Longleftrightarrow \frac{i}{\pi\alpha}
\sin\eta,
\label{eq:4}
\end{equation}
where $\eta$ represents the boson field and $\alpha$ is a short-range
cutoff of the order of the lattice spacing $a$. The corresponding
model is known in the literature as the phase Hamiltonian approach to
interacting electron-phonon systems.\cite{Fukuyama85} When the modulus
of the phonon field is frozen ($T\ll\Delta_0$), we are left with two
bosonic (phase) variables, $\eta,\phi$, whose dynamics is governed by
the Lagrangian\cite{Fukuyama85}
\begin{eqnarray}
{\cal L} & = & \frac{\hbar}{8\pi v_F} \left[ (\partial_{t}\eta)^2 -
v_F^2(\partial _{x}\eta)^2 \right] +
\frac{\Delta_0^2}{g^2\bar{\omega}^2}(\partial_{t} \varphi)^2 \nonumber
\\ & & + \frac{\varepsilon_F\Delta_0}{\pi\hbar v_F}\cos(\eta- \varphi)
+ \frac{2\Delta_0^2\omega_0^2} {g^2\bar{\omega}^{2}M^2}
\cos(\eta-\varphi + M\varphi).
\label{eq:5}
\end{eqnarray}
The last term in Eq.~(\ref{eq:5}) is the commensurability energy
written in the bosonized form,\cite{Lee74,Fukuyama85} where $\hbar
\omega_0 \simeq \hbar \bar{\omega}M (\Delta_0/ \varepsilon_F)^{M/2-1}$
($\omega_0 \ll \bar{\omega}$) is the so-called energy of the
commensurability ``pinning'' of the CDW.

The Lagrangian in Eq.~(\ref{eq:5}) is readily generalized to the case
of interacting electrons. For a forward scattering electron-electron
interaction one can replace the first two terms in Eq.~(\ref{eq:5}),
describing the free electron motion, by the analogous Lagrangian for a
LL,
\begin{equation}
{\cal L}_{LL} = \frac{\hbar}{8\pi K_\rho v_\rho} \left[
(\partial_{t}\eta)^2 -v_\rho^2(\partial_{x}\eta)^2 \right],
\label{eq:6}
\end{equation}
where $v_\rho$ is the charge velocity and $K_\rho$ is the bare
correlation parameter. These quantities can be expressed in terms of
the bare interaction constants (e.g., see Ref.~\onlinecite{Voit95}).

\begin{figure}
\setlength{\unitlength}{1mm}
\begin{picture}(80,70)(0,0)
\put(10,10){\epsfxsize=8cm\epsfbox{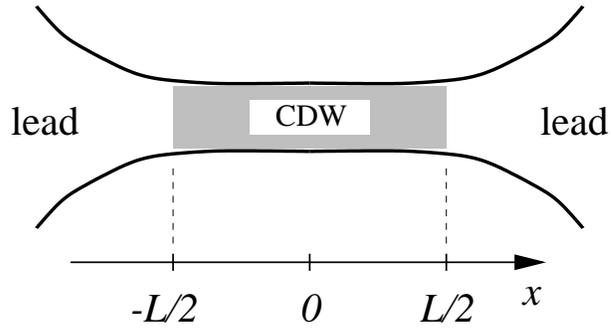}}
\end{picture}
\caption{A mesoscopic metal-CDW-metal junction.}
\label{fig1}
\end{figure}

We will use the Lagrangian density of Eq.~(\ref{eq:5}) to analyze the
electron transport through a finite CDW junction of the length $L$
(see Fig.~\ref{fig1}). Let us assume that the electron-phonon
interaction exists only in the region $-L/2<x<L/2$ and is
adiabatically switched off outside the junction. To model the
metal-CDW-metal junction one can multiply the last three terms in
Eq.~(\ref{eq:5}) by the step function $\Theta(x+L/2) \Theta(L/2-x)$,
confining the electron-phonon interaction to the finite region of
length $L$.

For a homogeneous system, the model represented by Eq.~(\ref{eq:5})
was studied in Ref.~\onlinecite{Fukuyama85} for two different cases:
(i) at a fixed phase of the phonon field, $\varphi = \mbox{const}$;
(ii) for an ``adiabatic'' CDW motion, when $\varphi(x,t) = \eta(x,t) +
\mbox{const}$. As we will see in what follows the last ansatz leads to
the correct quantum description of the CDW. Thus it can be called the
{\it quantum regime} of the CDW dynamics.

The physical motivation for studying the regimes mentioned above is
the following. Both cosine terms in Eq.~(\ref{eq:5}) describe the
backward scattering of electrons by the lattice distortions. Notice
that the term arising from the multiple Umklapp processes enters into
Eq.~(\ref{eq:5}) with a coefficient much smaller than that for the
direct electron-phonon interaction. Therefore, in the case of a fixed
phase [ansatz (i) above], one can neglect the effects of
commensurability and Eq.~(\ref{eq:5}) is reduced to the bosonic form
of a Lagrangian describing free massive Dirac fermions (with the gap
equal to $\Delta_0$). It gives us the Bogoliubov-de Gennes equations
for the electron wave function in the piecewise constant field of the
order parameter. So the ansatz (i) corresponds to the situation when
the dynamics of CDW is totally ignored. It describes only the
quasiparticle contribution to the conductivity which at temperatures
$T\ll\Delta_0\;$ is exponentially small and thus can be neglected.

At low temperatures one cannot neglect the quantum fluctuations of the
phonon field $\varphi(x,t)$ and a quantum description of the CDW
dynamics has to be adopted. The correct formulation assumes that the
electrons are in ``adiabatic motion'' \cite{Fukuyama85} with the
lattice distortion [ansatz (ii)]. In the Appendix we justify this
assumption by considering a simplified field-theoretical model for the
CDW. For the ansatz (ii) the most troublesome term ($\sim
\varepsilon_F$) in Eq.~(\ref{eq:5}), which arises from the direct
backscattering of electrons by phonons, disappears already at the
classical level. The resulting model is equivalent to the theory of a
commensurate CDW alone. For interacting electrons in Eq.~(\ref{eq:6}),
the corresponding Lagrangian takes the form
\begin{equation}
{\cal L}_{CDW} = \frac{\hbar}{8\pi K_{c}v_c} \left[
(\partial_{t}\varphi)^2- v_c^2(\partial_{x}\varphi)^2 +
\frac{2\Omega^2}{M^2}\cos M\varphi \right],
\label{eq:7}
\end{equation}
where
\begin{equation}
K_c = \frac{v_c}{v_\rho}K_\rho,
\label{eq:8}
\end{equation}
\begin{equation}
v_c = \frac{v_{\rho}}{\sqrt{1 + 8\pi
K_{\rho}v_{\rho}\Delta_0^2/g^2\bar{\omega}^2\hbar}},
\label{eq:9}
\end{equation}
and
\begin{equation}
\Omega^2 = \frac{8\pi K_c v_c}{\hbar} \left( \frac{\Delta_0\omega_0}
{g\bar{\omega}} \right)^2.
\label{eq:10}
\end{equation}
For noninteracting electrons ($K_\rho = 1$ and $v_\rho = v_F$),
Eq.~(\ref{eq:7}), taken in the limit $v_F \Delta_0^2 \gg
g^2\bar{\omega}^2\hbar$, coincides with the well-known Lagrangian for
a commensurate CDW. The corresponding parameters (CDW velocity and
pinning energy) are exactly the same ones found in the standard
approach by integrating out the fermionic degrees of freedom in
Eq.~(\ref{eq:1}),\cite{Krive86} namely,
\begin{equation}
v_c \simeq v_F \sqrt{\frac{g^2}{8\pi\hbar
v_F}}\left(\frac{\hbar\bar{\omega}} {\Delta_0}\right) \ll v_F \qquad
\mbox{with} \qquad \Omega \simeq \omega_0.
\label{eq:11}
\end{equation}

From a mathematical point of view, Eq.~(\ref{eq:7}) is a sine-Gordon
theory with well-known quantum properties. Notice that in the regime
of quantum transport our model [Eq.~(\ref{eq:7})], coincides (up to
the bare parameters $K_c$, $v_c$, and $\Omega$) with the Lagrangian
derived in Ref.~\onlinecite{Ponomarenko98b}. Therefore, the low
temperature properties of a CDW junction should be similar to those of
a correlated dielectric. Our analysis of the conductance of a CDW
junction can now be performed in analogy to that in
Ref.~\onlinecite{Ponomarenko98b}. Thus we will be brief in the
mathematical details and will concentrate on the physical
interpretations of our main results.

\section{The instanton contribution to the conductance}
\label{sec:III}

In an infinite system, the sine-Gordon model of Eq.~(\ref{eq:7}) is
characterized by an infinite set of equivalent vacua $|0\rangle_n =
|2\pi n/M\rangle$ ($n$ is integer). In a finite 1D system the ground
state is nondegenerate and if the length of the system is sufficiently
long, a unique vacuum can be found in the dilute instanton gas
approximation \cite{Rajaraman82}. The instanton approach to the
thermodynamics of finite CDW system was put forward for the first time
in Refs.~\onlinecite{Bogachek90} and \onlinecite{Krive92}, where the
partition function for a CDW ring threaded by a magnetic field was
calculated. Vacuum-to-vacuum tunneling introduces a new energy scale
\cite{Krive92}
\begin{equation}
\Delta_i \simeq \frac{\hbar v_c}{L} \sqrt{\frac{S_t^0}{2\pi\hbar}}
\exp \left(- \frac{S_t^0}{\hbar} \right)
\label{eq:12}
\end{equation}
($S_t^0$ is the
one-instanton tunnel action), which is nothing but the width of the
instanton band. In terms of the bare parameters shown in
Eqs.~(\ref{eq:7}), (\ref{eq:8}), (\ref{eq:9}), and (\ref{eq:10}) the
one-instanton action reads
\begin{equation}
S_t^0=\frac{E_s L}{\hbar v_c}=\frac{1}{K_c
M^2}\frac{2}{\pi}\frac{\hbar
\Omega}{\Delta_L^{\scriptsize{\mbox{CDW}}}},
\label{eq:13}
\end{equation}
where $E_s$ is the rest energy of the classical kink.

In the full quantum description the bare parameters of the sine-Gordon
model are renormalized by the interaction. For the gapped phase the
renormalized parameters are \cite{Luther74,Japaridze84} (see also the
review Ref.~\onlinecite{Kolomeisky96})
\begin{equation}
K_{c}^{\ast}M^2 = 1 \qquad \mbox{with} \qquad E_s\rightarrow
E_s^{\ast},
\label{eq:14}
\end{equation}
where $E_s^{\ast}$ is the energy of the quantum soliton. As it was
noticed in Ref.~\onlinecite{Ponomarenko98b}, this renormalization is
equivalent to the summation of all multi-loops contributions to the
renormalized one-instanton tunnel action $S_t^0\rightarrow S_t =
E_s^{\ast}/\Delta_L$. While the second prescription in
Eq.~(\ref{eq:14}) is physically evident, the first one needs some
comment. The simplest way to understand this result (so-called
Luther-Emery free fermion point \cite{Luther74}) is to consider the
situation when the chemical potential exceeds the gap. Let us at first
rescale the field $\varphi\rightarrow\varphi/M$ in
Eq.~(\ref{eq:7}). Then the same Lagrangian with the renormalized
parameters ($K_c\rightarrow K_c^{\ast}$ and $\Omega \rightarrow
\Omega^{\ast}$) has to describe the quantum phase of the sine-Gordon
model. In the case when chemical potential exceeds the gap
$E_s^{\ast}$, the system can be treated as a gas of weakly interacting
solitons at density $\rho_s$. It is then described by the first two
terms of the rescaled Lagrangian
\begin{equation}
{\cal L}_{\ast} = \frac{1}{8\pi v_c K_c^{\ast}(\rho_s)M^2} \left[
(\partial_{t} \varphi)^2 - v_c^2(\partial_{x}\varphi)^2 \right].
\label{eq:15}
\end{equation}
In the vicinity of a commensurate-incommensurate phase transition
($\rho_s \rightarrow 0$), the quantum solitons are noninteracting
particles. In the bosonic language, it means that $K_c^{\ast}(0)
M^2=1$ (see also Ref.~\onlinecite{Kolomeisky96}). The model in its
gapped phase, at this point, is equivalent to spinless massive Dirac
fermions.\cite{Coleman75,Kolomeisky96,Japaridze84}

The crucial difference between the closed CDW system studied in
Refs.~\onlinecite{Bogachek90} and \onlinecite{Krive92} and a finite
Peierls-Fr\"{o}hlich conductor connected to metallic leads is in the
boundary conditions imposed by the electron reservoirs on the CDW
dynamics. These boundary conditions can be derived by integrating out
electrons outside the CDW piece of the junction. As it was shown in
Ref.~\onlinecite{Ponomarenko98b}, this procedure results in a
Caldeira-Leggett (CL) action\cite{Caldeira81} for the boundary CDW
field $\varphi(x = \pm L/2,t)$. Physically, it means that electrons in
the leads induce friction (which appears as a logarithmic interaction
between instantons\cite{Caldeira81}) in the vacuum-to-vacuum
tunneling. Although, numerically, the logarithmic interaction only
changes the exponential prefactor in the tunnel energy shift
[Eq.~(\ref{eq:12})], it drastically affects the instantons
trajectories with a nonzero total topological charge. The action taken
on these paths diverges, and only trajectories with zero net charge
contribute to the partition function \cite{Schmid83}. This very
property allows one to use a dual representation
\cite{Schmid83,Ponomarenko98b} to studying transport properties of CDW
junction at low temperatures. In terms of the Josephson-like field
$\tilde{\Theta}(x,t)$, dual to the CDW field $\varphi(x,t)$, the
Lagrangian which yields the same partition function as that calculated
in the dilute instanton gas approximation takes the form (see
Ref.~\onlinecite{Ponomarenko98b})
\begin{eqnarray}
\tilde{\cal L} & = & \frac{\hbar}{8\pi v_c} \left[
(\partial_{t}\tilde{\Theta})^2 - v_c^2(\partial_{x}\tilde{\Theta})^2
\right] \nonumber \\ & & - \Delta_i^{(R)} \delta(x) \cos \left(
\tilde{\Theta}(x,t)/M \right).
\label{eq:16}
\end{eqnarray}
Here $\Delta_i^{(R)}$ is the instanton shift of the vacuum energy,
Eq.~(\ref{eq:12}), renormalized by the friction. The $1/M$ factor in
the last term is needed to take into account the fractional
topological charge $q_t=2\pi/M$ of individual instantons. The induced
CL-action does not introduce any new dimensional parameter to the
problem and affects only the prefactor in Eq.~(\ref{eq:12}) ( corrected
by the multiloop contributions, Eq~(\ref{eq:14})). Thus, up
to an irrelevant numerical factor, we have\cite{Ponomarenko98b}
$\Delta_i^{(R)} \simeq \Delta_i$ (friction, of course, could only
diminish the one-instanton vacuum energy shift).

After rescaling the dual field $\tilde{\Theta}/M \Longrightarrow
\Theta$, Eq.~(\ref{eq:16}) is transformed into the Lagrangian for a
quantum impurity problem in the infinite LL (the Kane-Fisher problem
\cite{Kane92}) with a correlation parameter $K_c = 1/M^2$. The desired
conduction in the initial problem (CDW- junction) is related to the
dimensionless conductance $g_{KF}$ of the dual problem by a simple
expression \cite{Schmid83,Ponomarenko98b}
\begin{equation}
G_{i}(T)=\frac{e^2}{h} \left[ 1-M^{2}g_{KF}(T) \right].
\label{eq:17}
\end{equation}
Notice the factor $M^2$ in Eq.~(\ref{eq:17}). In order to map the dual
model, Eq.~(\ref{eq:16}), onto the known problem, we rescaled the dual
field $\tilde\{\Theta(x,t)$. Since, in general, the conductance is
proportional to the square of the dynamical field, the dimensionless
conductance of the dual problem is $M^{2}g_{KF}$. Physically, the
additional factor of $M^2$ originates from the correct definition of
the Josephson-like current \cite{Ponomarenko98b} in the model of
Eq.~(\ref{eq:16}). Namely, it comes from the $M$-fold backscattering
current induced by the potential difference $MV$, where $V$ is the
voltage across the junction.

The quantity $g_{KF}(T)$ is known exactly (e.g., see
Ref.~\onlinecite{Fendley98}, where the current-voltage dependence for
the Kane-Fisher problem was obtained by using the dual transformation
method). For the CDW system the commensurability index $M$ is an
integer larger than 2 and the corresponding correlation parameter is
small, $K_c\ll 1$. In this case, the low- and high-temperature
asymptotics of the conductance take a simple form (up to irrelevant
numerical constants)
\begin{equation}
M^{2}g_{KF}(T,K_c\ll 1)\simeq\left\{ \begin{array}{ll}
\left(\frac{T}{\sqrt{K_c}\Delta_i}\right)^{2/K_c}, &
T\ll\sqrt{K_c}\Delta_i\\
1-\left(\frac{\sqrt{K_c}\Delta_i}{T}\right)^2, &
T\gg\sqrt{K_c}\Delta_i. \end{array} \right.
\label{eq:18}
\end{equation}
According to Eqs.~(\ref{eq:17}) and (\ref{eq:18}), the conductance of
a metal-CDW-metal junction is perfect at $T\rightarrow 0$. Loosely
speaking, very slow (low-energy) electrons from the metal leads, when
arriving at the CDW junction, see a strongly fluctuating,
translationally invariant electron-phonon system and, consequently,
are not backscattered by the lattice distortions. A significant
backscattering appears when the Heisenberg time $t_H \sim
\hbar/\varepsilon$ for quasiparticles coming from leads becomes
comparable or smaller than the characteristic lifetime of the
perturbative vacuum ($\sim \hbar/\Delta_i$). It is clear that at
$T>\Delta_i$ the instanton mechanism of charge transport predicts a
strongly suppressed conduction.

At finite temperatures, however, there is a mechanism competing with
the instanton contributions to charge transport in CDW systems. It is
associated with the thermally excited fractionally charged solitons of
the commensurate CDW.\cite{Gruner94,Krive86} The soliton contribution
to the persistent current in a CDW ring was considered in
Refs.~\onlinecite{Bogachek90} and \onlinecite{Krive92}. Thus, we now
proceed to calculate the soliton part, $G_s$, of the conductance of a
mesoscopic metal-CDW-metal junction.

\section{The soliton contribution to the conductance}
\label{sec:IV}

It is physically evident that at sufficiently ``high'' temperatures,
$T \gg \Delta_L$, the transport coefficients under consideration are
determined by the bulk properties of a Peierls- Fr\"ohlich system. In
the quantum regime of transport a commensurate CDW is described by a
quantum sine-Gordon model [the Lagrangian in Eq.~(\ref{eq:7}) at the
point $K_c^{\ast}M^2 = 1$]. It is known \cite{Luther74,Coleman75} that
this point corresponds to noninteracting Dirac fermions. Thus, at
temperatures $T \gg \Delta_L$ the problem of electron transport
through a CDW junction can be mapped onto the well-known problem of
transport of Dirac fermions. The latter in its turn is mathematically
equivalent to that of quasiclassical transport through a CDW
junction.\cite{Visscher96} The only important distinction is that in
our case the electric charge of the Dirac fermions is fractional
$q_s=e/M$ (they are solitons of the quantum sine-Gordon model).

The soliton contribution to the conductance can be calculated using
Landauer formula,\cite{Landauer}
\begin{equation}
G_s(T) = \frac{q_s^2}{h} \int_{-\infty}^{\infty} d\varepsilon \left(
-\frac{\partial f_{FD}} {\partial\varepsilon}\right)
T_t(\varepsilon),
\label{eq:19}
\end{equation}
where $f_{FD}$ is the Fermi-Dirac distribution function and
$T_t(\varepsilon)$ is the transmission probability of massless
fermions through a ``gapped'' region. The transmission coefficient can
be found by matching the wave functions at the boundary points $x=\pm
L/2$. The result is (see also Ref.~\onlinecite{Ponomarenko98a})
\begin{equation}
T_t(\varepsilon) = \frac{\varepsilon^2-E_s^{\ast 2}} {\varepsilon^2
-E_s^{\ast 2} + E_s^{\ast 2} \sin^2 \left(
\sqrt{\varepsilon^2-E_s^{\ast 2}}/ \Delta_L \right)}.
\label{eq:20}
\end{equation}
By carrying out the integral in Eq.~(\ref{eq:19}), it is easy to find
the following low-$T$ ($T\ll E_s^{\ast}$) and high-$T$ asymptotes
\begin{equation}
G_s(T) \simeq \left\{ \begin{array}{ll} \frac{e^{2}4}{hM^2}
\exp(-\frac{2E_s^{\ast}}{\Delta_L}), & \Delta_L\ll T\ll E_s^{\ast} \\
\frac{e^2}{hM^2} \left( 1-2\pi\frac{E_s^{\ast 2}}{\Delta_L T} \right),
& T\gg  E_s^{\ast 2}/\Delta_L
\end{array}.
\right.
\label{eq:21}
\end{equation}
As one can see from Eq.~(\ref{eq:21}), the soliton conductance at low
temperatures is exponentially small and temperature independent. It
corresponds to the tunneling of fractionally charged particles through
a dielectric region. At high temperatures ($T > E_s^{\ast}$), the
thermally excited solitons and antisolitons yield a perfect (in terms
of the fractional charge $q_s$) conductance $G_M = q_s^{2}/h$.

\begin{figure}
\setlength{\unitlength}{1mm}
\begin{picture}(100,80)(0,0)
\put(10,10){\epsfxsize=10cm\epsfbox{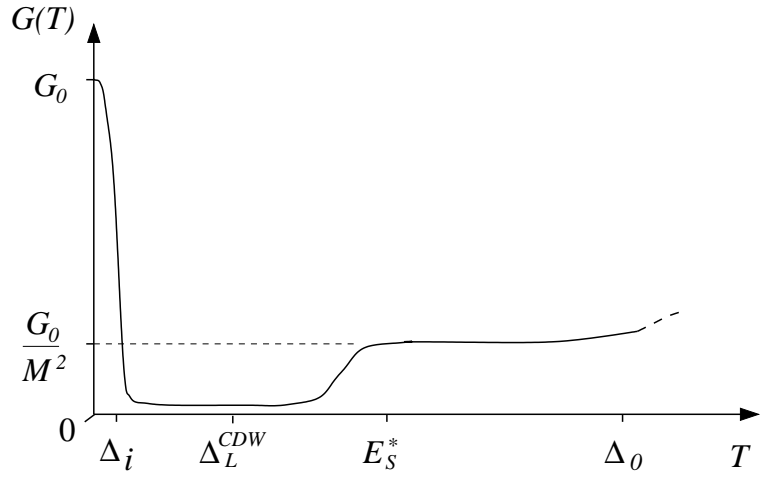}}
\end{picture}
\caption{Conductance as function of temperature.}
\label{fig2}
\end{figure}

The total conductance can be represented by the interpolative formula
$G(T) \simeq G_i(T) + G_s(T)$. This is schematically shown on
Fig.~\ref{fig2}. From Eqs.~(\ref{eq:17}) and (\ref{eq:18}) one can
readily find that the ``instanton'' part of the conductance matches
the soliton contribution, Eq.~(\ref{eq:21}), at temperatures $T_m \sim
\sqrt{\Delta_L E_s^{\ast}} \gg \Delta_L$. As a result, in a wide
temperature interval $T_m <T< E_s^{\ast}$ the conductance is
exponentially small and practically temperature independent, namely,
$G_m \sim G_0 \exp(-2E_s^{\ast}/\Delta_L)$. This trough-like shape of
the $G \times T$ curve changes at $T \sim E_s^{\ast}$ to the step-like
form, with $G \approx G_M = G_0/M^2$ which characterizes the transport
of fractional charge along the CDW junction.

\section{Final remarks}
\label{sec:V}

It is evident from the above considerations that in the case of an
incommensurate CDW, where the last term in Eq.~(\ref{eq:5}) is absent,
the temperature dependence of the dc conductance is much simpler than
for a commensurate CDW. The quantum theory of an incommensurate CDW is
equivalent in many aspects to a theory of a Luttinger liquid. So, for
adiabatic contacts the conductance of an incommensurate CDW junction
equals $G_0$ and it is temperature independent. For a commensurate CDW
the step in the conductance which corresponds to the soliton mechanism
of conductivity will last (for a purely 1D system) up to temperatures
of order $\Delta_0$. At this point, the conductance begins to increase
and eventually reaches the pure metallic value, $e^2/h$, due to the
contribution of thermally excited quasiparticles (electrons and
holes). Moreover, the temperature suppression of the gap $\Delta(T)$
in the quasiparticle spectrum also becomes important. For quasi-1D
systems, the restoration of metallic conductivity will happen, of
course, at much low temperatures, in the vicinity of the Peierls phase
transition.

The above consideration holds for the case when electron-phonon
interaction in 1D system is strong enough to produce Peierls
gap.\cite{Voit87} Otherwise, the interacting electron system falls
into the Luttinger liquid class of universality with parameters
determined both by electron-electron and electron-phonon interactions
(see Refs.~\onlinecite{Voit87} and \onlinecite{Shizuya}).


\acknowledgements

This work was partially supported by the NRC Grant for Twinning
Program between USA and Ukraine and by the Brazilian agency
CNPq. I.V.K. thanks E. Bogachek, L. Gorelik, U. Landman, A. Nersesyan,
and R. Shekhter for fruitful discussions. I.V.K. is also greatful to
the School of Physics at the Georgia Institute of Technology for
hospitality. E.R.M. thanks P. Lee for critical comments.


\begin{appendix}

\section{}

Using a simpler version of the Lagrangian of Eq.~(\ref{eq:5}), we will
argue in favor of the ansatz (ii) of Sec.~\ref{sec:II}. Let us thus
begin with the following Lagrangian, describing a relativistic,
bosonized electron-phonon system of the incommensurate type,
\begin{eqnarray}
{\cal L} & = & \frac{1}{2} \partial_\mu \eta \partial^\mu \eta +
\frac{\rho_0^2}{2} \partial_\mu \phi \partial^\mu \phi + A\rho_0 \cos
\beta (\eta - \phi) \nonumber \\ & & - \frac{\beta}{2\pi} \partial_\mu
\eta \epsilon^{\mu\nu} a_\nu,
\end{eqnarray}
where $\eta$ is the bosonic field related to the electronic degrees of
freedom, while $\phi$ and $\rho_0$ represent the phase and amplitude
of the phonon field (we only allow for phase fluctuations). The gauge
field $a_\nu$ is the external source of electric field. Already at the
classical level we can see that there exist two distinct modes in the
system. Setting the electric field to zero, we find the equations of
motion
\begin{equation}
\partial_\mu \partial^\mu (\eta + \rho_0^2 \phi) = 0
\label{eq:massless}
\end{equation}
and
\begin{equation}
\partial_\mu \partial^\mu (\eta - \phi) + A\beta \left( \rho_0 +
\frac{1}{\rho_0} \right) \sin \beta (\eta - \phi) = 0.
\label{eq:massive}
\end{equation}
Equation (\ref{eq:massless}) describes a massless mode, whereas
(\ref{eq:massive}), a sine-Gordon equation, gives rise to a massive
mode. We expect the low-energy physics to be controlled by the
former. Indeed, the electrical conductance of this system can be
obtained from the expectation value of the current
\begin{equation}
\langle J^\nu \rangle = -i \left. \frac{\partial \ln Z}{\partial
a_\nu} \right|_{a=0},
\end{equation}
where the partition function is given by
\begin{equation}
Z = \int D\eta D\phi\ e^{-i\int d^2 x {\cal L}}.
\end{equation}
Introducing ``relative'' and center-of-mass'' fields
\begin{equation}
\chi \equiv \eta - \phi
\end{equation}
and
\begin{equation}
\xi \equiv \frac{\phi + \rho_0^2 \phi}{1 + \rho_0^2},
\end{equation}
we can write this partition function as $Z[a] =
Z_{\scriptsize{\mbox{rel}}}[a] Z_{\scriptsize{\mbox{CM}}} [a]$, where
\begin{equation}
Z_{\scriptsize{\mbox{rel}}} = \int D\chi\ e^{ -i \int d^2x {\cal
L}_{\scriptsize{\mbox{rel}}}}
\end{equation}
with
\begin{eqnarray}
{\cal L}_{\scriptsize{\mbox{rel}}} & = &
\frac{\rho_0^2}{2(1+\rho_0^2)} \partial_\mu \chi \partial^\mu \chi + A
\rho_0 \cos \beta \chi \nonumber \\ & & - \frac{\beta
\rho_0^2}{2\pi(1+\rho_0^2)} \partial_\mu \chi \epsilon^{\mu\nu} a_\nu
\end{eqnarray}
and
\begin{equation}
Z_{\scriptsize{\mbox{CM}}} = \int D\xi\ e^{ -i \int d^2x {\cal
L}_{\scriptsize{\mbox{CM}}}}
\label{eq:zxi}
\end{equation}
with
\begin{equation}
{\cal L}_{\scriptsize{\mbox{CM}}} = \frac{1+\rho_0^2}{2} \partial_\mu
\xi \partial^\mu \xi - \frac{\beta}{2\pi} \partial_\mu \xi
\epsilon^{\mu\nu} a_\nu.
\end{equation}
The path integration in Eq.~(\ref{eq:zxi}) can be carried out exactly,
since the action is quadratic. The modes described by the Lagrangian
${\cal L}_{\scriptsize{\mbox{CM}}}$ are massless (gapless). On the
other hand, the modes in ${\cal L}_{\scriptsize{\mbox{rel}}}$ are
massive due to the presence of the cosine term. The contribution of
these modes to the current is strongly suppressed. At low
temperatures, the massive modes can be neglected all together in
comparison to the massless ones. This, in turn, is equivalent to
setting $\chi= \mbox{const.}$ in the theory, in accordance to the
ansatz (ii) of Sec.~\ref{sec:II}. As a result, the current is solely
determined by the center-of-mass motion,
\begin{equation}
\langle J^\nu \rangle = -i \left. \frac{\partial \ln
Z_{\scriptsize{\mbox{CM}}}}{\partial a_\nu} \right|_{a=0}.
\end{equation}

\end{appendix}


\end{multicols}


\end{document}